\documentstyle[11pt,aaspp4]{article}
\input psfig.tex

\def\gsim{\;\rlap{\lower 2.5pt
 \hbox{$\sim$}}\raise 1.5pt\hbox{$>$}\;}
\def\lsim{\;\rlap{\lower 2.5pt
   \hbox{$\sim$}}\raise 1.5pt\hbox{$<$}\;}
\newcommand\beq{\begin{equation}}
\newcommand\eeq{\end{equation}}

\def\v{\vspace{-0.1in}}

\begin{document}

\Large
\centerline{\bf Gamma-ray background from structure }
\centerline{\bf formation in the intergalactic medium} 
\normalsize
\author{\bf Abraham Loeb$^{\star}$ and Eli
Waxman$^{\dagger}$}
\medskip
\noindent
$\star$ Harvard-Smithsonian CfA, 60 Garden Street,
Cambridge, MA 02138, USA\\ 
\noindent
$\dagger$ Department of Condensed Matter Physics,
Weizmann Institute, Rehovot, 76100, Israel

\bigskip
\centerline{\bf To appear in {\it Nature}. (Under press embargo until
published.)}

\vskip 0.2in 
\hrule 
\vskip 0.2in 

{\bf The universe is filled with a diffuse and isotropic extragalactic
background of $\gamma$-ray radiation$^1$, containing roughly equal energy
flux per decade in photon energy between 3 MeV--100 GeV. The origin of this
background is one of the unsolved puzzles in cosmology. Less than a quarter
of the $\gamma$-ray flux can be attributed to unresolved discrete
sources$^{2,3}$, but the remainder appears to constitute a truly diffuse
background whose origin has hitherto been mysterious.  Here we show that
the shock waves induced by gravity during the formation of large-scale
structure in the intergalactic medium, produce a population of
highly-relativistic electrons with a maximum Lorentz factor $\ga 10^7$.
These electrons scatter a small fraction of the microwave background
photons in the present-day universe up to $\gamma$-ray energies, thereby
providing the $\gamma$-ray background.  The predicted diffuse flux agrees
with the observed background over more than four decades in photon energy,
and implies a mean cosmological density of baryons which is consistent with
Big-Bang nucleosynthesis.  }

The universe started from a smooth initial state, with small density
fluctuations that grew in time due to the effect of gravity. The formation
of structure resulted in non-relativistic, collisionless, shock waves in
the baryonic gas due to converging bulk flows, and raised its mean
temperature to $\sim 10^7$ K ($=1$ keV) at the present time$^4$. This is
indeed the characteristic temperature of the warm gas in groups of
galaxies, which are the typical objects forming at the present epoch. The
intergalactic shocks are usually strong, since the gravitationally-induced
bulk flows are often characterized by a high Mach number.  The
intergalactic gas may have also been heated by shocks driven by 
outflows from young galaxies$^{5-7}$.

Collisionless, non-relativistic, shocks are known to generically accelerate
a power-law distribution of relativistic electrons, with a number density
per electron momentum, $p_e$, of $dn_e/dp_e= Kp_e^{-\alpha}$, where $K$ is
a constant$^8$.  The power-law index for strong shocks in a gas with an
adiabatic index $\Gamma=5/3$, is$^{8-10}$, $\alpha=[(r+2)/(r-1)]=2$, where
$r=[(\Gamma+1)/(\Gamma-1)]$ is the shock compression factor. Such an
electron distribution is found in the strong shocks surrounding supernova
remnants in the interstellar medium$^{8}$.  Recent X-ray$^{11,12}$ and
TeV$^{13,14}$ observations of the supernova remnants SN1006 and SNR RX
J1713.7--3946 imply that electrons are accelerated in the remnant shocks up
to an energy $\sim 100$~TeV, and are confined to the collisionless fluid by
magnetic fields.
These shocks have a velocity of order $10^3~{\rm km~s^{-1}}$, similar to
the velocity of the intergalactic shocks we consider here.  The inferred
energy density in relativistic electrons constitutes $1$--$10\%$ of the
post-shock energy density in these remnants, a fraction consistent with the
global ratio between the mean energy density of cosmic-ray electrons and
the turbulent energy density in the interstellar medium of our galaxy.

Since the physics of shock acceleration can be scaled up to intergalactic
distances, it appears natural to assume that a similar population of
relativistic electrons was also produced in the intergalactic medium.  The
maximum Lorentz factor of the accelerated electrons, $\gamma_{\rm max}$, is
set by equating their acceleration and cooling times. The $e$-folding time
for the acceleration of the relativistic electrons is $t_{\rm acc}\sim
(r_{\rm L}c/v_{\rm sh}^2)$, where $v_{\rm sh}=
(\Gamma+1)[kT/(\Gamma-1)m_p]^{1/2}$ is the shock velocity relative to the
unshocked gas, $m_p$ is the proton mass, and $r_{\rm L}$ is the electron
Larmor gyration radius. For an electron with a Lorentz factor $\gamma_e$,
$r_{\rm L}= 5.5\times 10^{-2} (\gamma_7/B_{-7})$ pc, where
$\gamma_{7}=(\gamma_e/10^7)$, and $B_{-7}$ is the magnetic field strength
in units of $0.1\mu$G. A magnetic field amplitude of $0.1$--$1 \mu$G is
often detected in the halos of galaxy clusters$^{15-17}$.  In particular, a
magnetic field amplitude of $\ga 0.1 \mu$G was inferred on the multi-Mpc
scale of the Coma--A1367 supercluster$^{18}$, which provides a good example
for the intergalactic structures of interest here.  For such magnetic field
amplitude we get, $t_{\rm acc}\sim 2\times 10^4~{\rm
yr}~(\gamma_7/B_{-7})(kT/{\rm keV})^{-1}$.

The acceleration time is much shorter than the lifetime of the
intergalactic shocks, which is comparable to the age of the universe.  The
maximum Lorentz factor of the electrons, $\gamma_{\rm max}$, is therefore
not limited by the lifetime of their accelerator, but rather by their
cooling, primarily due to Compton scattering off the cosmic microwave
background$^{19}$.  Synchrotron cooling is negligible for $B_{-7}\la 10$.
The characteristic cooling time due to inverse-Compton scattering is
$t_{\rm cool}=[m_e c/ (4/3) \sigma_T \gamma_e u_{\rm cmb}] =1.2 \times
10^{10}~{\rm yr}\left({\gamma_e/200}\right)^{-1}$, where $m_e$ is the
electron mass, $\sigma_T$ is the Thomson cross-section, and $u_{\rm cmb}$
is the energy density of the cosmic microwave background.  By equating the
acceleration and cooling times, $t_{\rm acc}=t_{\rm cool}$, we find
$\gamma_{\rm max}= 3.7\times 10^7[B_{-7}(kT/{\rm keV})]^{1/2}$.

The estimates given above are valid as long as the electron Larmor radius
is smaller than the coherence length of the magnetic field.  Since the
Larmor radius of the relativistic electrons is extremely small, $r_{\rm
L}\sim0.1{\rm\,pc}$, this condition is satisfied even if the field
coherence length is much shorter than the $\sim$ Mpc scale of the
intergalactic shocks.

All electrons with $\gamma_e>200$ lose their energy over the age of the
universe. The energy of these electrons is converted to a diffuse
background of photons, which are produced through inverse-Compton
scattering of microwave background photons. The initial energy of a
microwave photon, $h\nu_0$, is boosted by the scattering up to an average
value of $h\nu= (4/3) \gamma_e^2 h\nu_0$.  Substituting the mean frequency
of the microwave background photons for $\nu_0$, we find that the
accelerated intergalactic electrons produce a diffuse background of
radiation extending from the UV [$h\nu= 36 (\gamma_e/200)^2$~eV] and up to
extreme $\gamma$-ray energies [$h\nu= 89 \gamma_{7}^2$~GeV]. For a
power-law distribution of relativistic electrons, $dn_e/dp_e=K p_e^{-2}$,
the energy density of the scattered radiation is predicted to be constant
per logarithmic frequency intervals, $\nu(du/d\nu)=Kc/2=u_e/2\ln\gamma_{\rm
max}$.  Here $u_e=Kc\ln\gamma_{\rm max}$~~is the energy density in
relativistic electrons produced by the intergalactic shocks.  If a fraction
$f_{\rm sh}$ of the baryons in the universe were shocked to a
(mass-weighted) mean temperature $T$, and a fraction $\xi_e$ of the shock
thermal energy was transferred to relativistic electrons, then
$u_e=\frac{3}{2}\xi_e f_{\rm sh} n_p kT= 5.1 \times 10^{-16} \xi_e f_{\rm
sh} \left({\Omega_b h_{70}^2/0.04}\right) \left(kT/{\rm keV}\right) ~{\rm
erg~cm^{-3}}$, where $n_p$ is the average proton density, $\Omega_b$ is the
cosmological baryon density parameter, and $h_{70}$ is the Hubble constant
in units of $70~{\rm km~s^{-1}~Mpc^{-1}}$.  By substituting $\ln\gamma_{\rm
max}=\ln (4\times 10^7)$, we then get
\begin{equation}
E^2{dJ\over dE}= 1.1 \left({\xi_e\over 0.05}\right)
\left({\Omega_bh_{70}^2\over 0.04}\right) \left(f_{\rm sh} kT\over {\rm
keV}\right)~~~{\rm keV~cm^{-2}~s^{-1}~sr^{-1}},
\label{eq:flux}
\end{equation}
where $E=h\nu$, and $J$ is the number flux of photons per solid angle.
Adopting the result from hydrodynamic simulations of$^4$ $f_{\rm sh}
(kT/{\rm keV})\sim 1$, the Big-Bang nucleosynthesis value
of$^{20,21}$~~$\Omega_{b}h_{70}^2=0.04^{+.014}_{-.027}$ (based on the
deuterium abundance), and the value of $\xi_e\sim 0.05$ inferred from
non-relativistic collisionless shocks in the interstellar medium, we obtain
a background flux of $\sim1~{\rm keV~cm^{-2}~s^{-1}~sr^{-1}}$ in the photon
energy range of 3 MeV to 100 GeV. As Figure 1 illustrates, this result is
in excellent agreement with the $\gamma$-ray background detected by the
EGRET instrument aboard the CGRO satellite$^1$.  

The sky-averaged contribution of identified extragalactic sources, such as
blazars ($\gamma$-ray loud active-galactic-nuclei), amounts to only $\la
7\%$ of the extragalactic $\gamma$-ray flux$^{3}$.  As shown in Figure 1,
the recent EGRET data is consistent with a luminosity function of faint,
undetected blazars, which could account for only $\la 25\%$ of the
unresolved $\gamma$-ray background$^{2,3}$.  The remainder of the hard
$\gamma$-ray background is expected to be highly isotropic in our model
because it is integrated over a large volume throughout the universe.  Its
large-angle fluctuations should be comparable (to within a factor of a few)
to that of the hard X-ray background, of which$^{25}$ $\sim 40\%$ is
contributed by early-type galaxies at $z\la 1$, since these galaxies trace
the same large scale structure as painted by the comic web of intergalactic
shocks.  This implies a root-mean-square fluctuation amplitude smaller than
$\sim 5\%$ on angular scales larger than a degree$^{26}$.  The predicted
high level of isotropy can serve as a test of our model, as soon as the
isotropy of the $\gamma$-ray background is measured to a higher precision
than presently available.  Another possible test is the spectral distortion
induced in the microwave background band by the low-energy tail of the
non-thermal intergalactic electrons. This distortion is, however, well
below the upper limit inferred from the COBE satellite data$^{27}$.

At photon energies $\la$ MeV the scattered background is sub-dominant
relative to the cumulative flux produced by discrete sources, such as
active galactic nuclei. In particular, the predicted scattered flux amounts
only to $\sim 10$--$15\%$ of the soft X-ray background measured by ROSAT at
1--2 keV, and is consistent with the upper limit of $\sim 20$--$30\%$ on
its unresolved fraction$^{25,28}$.

Most of the $\gamma$-ray background is produced during the latest episode
of shock heating of the intergalactic gas at $z\la 1$, since the mean gas
temperature is expected to decrease rapidly with increasing redshift$^4$.
Furthermore, the photons emitted at redshifts $z\ga 0.5$ lose a significant
fraction of their energy due to the expansion of the universe.  A more
accurate calculation of the $\gamma$-ray background can be achieved by
simulating the hydrodynamic evolution of the intergalactic medium and
injecting a population of accelerated electrons into each fluid element
which passes through a shock, at all redshifts. Such a simulation can yield
both the background flux and its statistical fluctuations on the sky. For a
given cosmological model of structure formation, the derived background
flux will be proportional to $\xi_e$, and will not depend on any other free
parameters.

Although the warm ($\la 10^7~$K) intergalactic medium is expected
theoretically to include a major fraction of the baryons in the present-day
universe, it has not been observed directly as of yet. The soft X-ray
emission from intergalactic structures, such as sheets and filaments or
galaxy groups, is often too faint to be detectable with current
telescopes$^4$. The observed cosmological density of stars$^{21}$,
$\Omega_{\star}h_{70}=3.5\times 10^{-3}$, is an order of magnitude smaller
than the baryon density predicted by Big-Bang nucleosynthesis.  If the
$\gamma$-ray background indeed results from keV shocks in the intergalactic
medium, then we can derive a lower limit on the mean baryon density
required to produce its observed flux. This minimum is obtained by
substituting $\xi_e=1/2$ (equipartition) and $f_{\rm sh}=1$ in equation
(\ref{eq:flux}), yielding $\Omega_{b} h_{70}^2\ga 4\times 10^{-3}$.  The
inferred lower limit is {\it larger} than the observed mass density in
stars.  Thus, if the reality of our model is proven, then this would not
only explain the origin of the diffuse component of the $\gamma$-ray
background, but also provide the first evidence for the ``missing baryons''
(note that even if the truly diffuse component amounts to only $\sim20\%$
of the $\gamma$-ray background flux, the more plausible range of $\xi_e\la
10\%$ yields a lower limit on the baryon density which is still larger than
the mass density in stars).

In our model, the $\gamma$-ray background is produced in the dense
filaments and sheets which channel gas from converging bulk flows in the
intergalactic medium. The densest and hottest shocks occur at the
intersections of these filaments around the locations of clusters of
galaxies. Although the richest accreting clusters may be rare and
contribute a small fraction of the background, they contain the brightest
shocks in the sky and produce the strongest fluctuations in the diffuse
$\gamma$-ray background. Direct detection of these shocks can be used to
calibrate $\xi_e$ in our model. Even a statistical detection of a
cross-correlation signal between background fluctuations and other sources
that trace the same large scale structure, such as galaxies, X-ray gas, the
Sunyaev-Zel'dovich effect, or synchrotron emission from the high energy
electrons in the intergalactic medium, can be used to prove the reality of
our model.  We predict that as cold gas goes through the strong
virialization shock of an accreting cluster, it emits non-thermal radiation
with photon energies between 30 eV and TeV. The shocked gas loses a
fraction $\sim (4.5/7)\times \xi_e\sim3\%$ of its thermal energy in this
process.  For $h\nu \gg 10$ keV, the cooling time of the corresponding
relativistic electrons is shorter than a billion years.  In this regime the
total non-thermal luminosity of the cluster, $L_{\rm nt}$, is limited by
the time it takes the cluster gas to cross the virialization shock,
i.e. $t_{\rm vir}\sim (2~{\rm Mpc}/2\times 10^3~{\rm km~s^{-1}})= 10^{9}$
yr. For a young cluster which shocks gas of total mass $M_{\rm gas}$ to a
virial temperature $T_{\rm gas}$, the total (bolometric) non-thermal
luminosity is
\begin{equation}
L_{\rm nt}\sim \left({4.5\xi_e/7 \over t_{\rm vir}}\right)\left(M_{\rm
gas}\over m_p\right) \left({3\over 2} kT_{\rm gas}\right) = 1.5\times
10^{45} \left({\xi_e\over 0.05}\right) \left({10^9~{\rm yr}\over t_{\rm
vir}}\right) \left({M_{\rm gas}\over 10^{14} M_\odot}\right) \left({kT_{\rm
gas}\over 5~{\rm keV}}\right)~{\rm erg~s^{-1}}.
\label{eq:cluster_luminosity}
\end{equation}
(The total cluster mass, including the dark matter, is typically an order
of magnitude larger than $M_{\rm gas}$.)  The luminosity per each
logarithmic interval in photon energy, anywhere between 30 eV and TeV, is
$\sim 0.1L_{\rm nt}$.  This emission component would affect estimates of
the gas temperature and the thermal luminosity of young clusters which form
through shocking of cold gas.  However, the high-energy emission would be
suppressed in the weak shocks that result from mergers of pre-existing
clusters which contain pre-heated gas, since the energy distribution of the
accelerated electrons is steeper for weak shocks$^8$.  As a cluster relaxes
to hydrostatic equilibrium and ages, its non-thermal luminosity declines.
The emission of hard radiation disappears almost entirely as soon as there
is no strong shocking of gas.  At a perfectly quiescent state of
hydrostatic equilibrium, the non-thermal radiation spectrum cuts-off above
a photon energy, $h\nu \sim 0.5~{\rm keV}(\tau/3\times 10^{9}~{\rm
yr})^{-2}$, for which the electron cooling time equals the cluster age,
$\tau$.  Detection of this cut-off can be used as a method of dating the
time that has passed since the last major accretion episode of cold gas by
the cluster.  Most relaxed clusters with ages $\tau\gg 10^9$ yr, are
expected to show only weak non-thermal emission at $h\nu \ga 10$ keV.  We
note that the detection of nonthermal X-ray emission has been reported
recently for several clusters$^{16,17,29}$.  In addition, some X-ray
clusters possess radio halos$^{30}$ which may result from synchrotron
emission by the residual population of relativistic electrons in the
cluster magnetic field.

\vskip 0.45in
\hrule
\vskip 0.15in
\small
\noindent
\begin{itemize}
\item[1.]  Sreekumar, P. {\it et al.} EGRET observations of the
extragalactic gamma-ray emission. {\it Astrophys. J.} {\bf 494}, 523-534
(1998).  
\v
\item[2.] Chiang, J. \& Mukherjee, R. The luminosity function of the EGRET
gamma-ray blazars. {\it Astrophys. J.} {\bf 496}, 752-760 (1998).
\v
\item[3.] Mukherjee, R. \& Chiang, J. EGRET $\gamma$-ray blazars:
luminosity function and contribution to the extragalactic $\gamma$-ray
background. {\it Astroparticle Phys.} {\bf 11}, 213-215 (1999).  
\v
\item[4.] Cen, R. \& Ostriker, J. P. Where are the baryons?. {\it
Astrophys. J.}  {\bf 514}, 1-6 (1999).
\v
\item[5.] Metzler, C. A. \& Evrard, A. E. A simulation of the intracluster
medium with feedback from cluster galaxies. {\it Astrophys. J.} {\bf 437},
564-583 (1994).
\v
\item[6.] Nevalainen, J., Markevitch, M. \& Forman, W. R.  The cluster M-T
relation from temperature profiles observed with ASCA and ROSAT. {\it
Astrophys. J.}  {\bf in press} (2000); astro-ph/9911369.
\v
\item[7.] Loewenstein, M. Heating of intergalactic gas and cluster scaling
relations.  {\it Astrophys. J.}  {\bf in press} (2000); astro-ph/9910276.
\item[8.] Blandford, R. \& Eichler, D. Particle acceleration at
astrophysical shocks -- a theory of cosmic-ray origin. {\it Phys. Reports}
{\bf 154}, 1-75 (1987).
%\noindent
%5. Axford, W. I., Leer, E., \& Skadron, G. {\it Proc. 15th Int. Conf. Cosmic
%Rays}, Plovdiv, Bulgaria (1977).\\
\v
\item[9.] Bell, A. R. The acceleration of cosmic rays in shock fronts. II.
{\it Mon. Not. R. astr. Soc.} {\bf 182}, 147-156 (1978).
\v
\item[10.] Blandford, R. D., \& Ostriker, J. P. Particle acceleration by
astrophysical shocks. {\it Astrophys. J.} {\bf 221}, L29-L32 (1978).
\v
\item[11.] Koyama, K. {\it et al.} Evidence for shock acceleration of
high-energy electrons in the supernova remnant SN:1006. {\it Nature} {\bf
378}, 255-258 (1995).
\v
\item[12.] Koyama, K. {\it et al.} Discovery of non-thermal x-rays from the
northwest shell of the new SNR RX J1713.7-3946: The Second SN 1006?. {\it
PSAJ} {\bf 49}, L7-L11 (1997).
\v
\item[13.] Tanimori, T. {\it et al.} Discovery of TeV gamma rays from SN
1006: further evidence for the supernova remnant origin of cosmic
rays. {\it Astrophys. J.} {\bf 497}, L25-L28 (1998).
\v
\item[14.] Muraishi, {\it et al.} Evidence for TeV gamma-ray emission from
the shell type SNR RXJ1713.7-3946. {\it Astron. \& Astrophys.}, {\bf in
press} (2000); astro-ph/0001047.
\v
\item[15.] Kronberg, P. Extragalactic magnetic fields.  {\it
Rep. Prog. Phys.} {\bf 57}, 325-382 (1994).
\v
\item[16.] Fusco-Femiano, R. {\it et al.} Hard x-ray radiation in the Coma
cluster spectrum. {\it Astrophys. J.} {\bf 513}, L21-L24 (1999).
\v
\item[17.] Rephaeli, Y., Gruber, D. \& Blanco, P.~ Rossi X-Ray Timing
Explorer observations of the Coma cluster. {\it Astrophys. J.} {\bf 511},
L21-L24 (1999).
\v
\item[18.] Kim, K.-T., Kronberg, P. P., Giovannini, G. \& Venturi, T.
Discovery of intergalactic radio emission in the Coma-A1367
supercluster. {\it Nature} {\bf 341}, 720-723 (1989).
%\noindent
%14. Dreher, J. W., Carilli, C. L. \& Perley, R. A. {\it Astrophys. J.}
%{\bf 316}, 611-625 (1987).\\
%\noindent
%15. Taylor, G. B. \& Perley, R. A. {\it Astrophys. J.}
%{\bf 416}, 554-562 (1993).\\
%\noindent
%16. Ge, J. P. \& Owen, F. N {\it Astron. J.} {\bf 105}, 778-787 (1993);
%{ibid} {\bf 108}, 1523-1533 (1994).\\
\v
\item[19.] Felten, J. E., \& Morrison, P. Omnidirectional inverse Compton
and synchrotron radiation from cosmic distribution of fast electrons and
thermal photons. {\it Astrophys. J.} {\bf 146}, 686-708 (1966).
\v
\item[20.] Tytler, D., O'Meara, J. M., Suzuki, N., \& Lubin, D.  Review of
Big Bang nucleosynthesis and primordial abundances.  {\it Physica Scripta}
{\bf in press} (2000); astro-ph/0001318.
\v
\item[21.] Fukugita, M., Hogan, C. J. \& Peebles, P. J. E. The cosmic
baryon budget. {\it Astrophys. J.}  {\bf 503}, 518-530 (1998).
\v
\item[22.] Primack, J. R., Bullock, J. S., Somerville, R. S. \& McMinn, D.
Probing galaxy formation with TeV gamma ray absorption. {\it Astroparticle
Physics} {\bf 11}, 93-102 (1999).
\v
\item[23.] Konopelko, A. K., Kirk, J. G., Stecker, F. W. \& Mastichiadis,
A.  Evidence for intergalactic absorption in the TeV gamma-ray spectrum
of Markarian 501. {\it Astrophys. J.} {\bf 518}, L13-L15 (1999).  
\v
\item[24.] Coppi, P. S. \& Aharonian, F. A.  Constraints on the very high
energy emissivity of the universe from the diffuse GeV gamma-ray
background. {\it Astrophys. J.} {\bf 487}, L9-L12 (1997).
\v
\item[25.] Mushotzky, R. F., Cowie, L. L., Barger, A. J. \& Arnaud,
K. A. Resolving the extragalactic hard X-ray background. {\it Nature} {\bf
in press} (2000); astro-ph/0002313.
\v
\item[26.] Fabian, A. C. \& Barcons, X. The origin of the X-ray
background. {\it Ann. Rev. of Astron. \& Astrophys.} {\bf 30}, 429-456
(1992).  
\v
\item[27.] Fixsen, D. J. {\it et al.}
%, Cheng, E. S., Gales, J. M., Mather, J. C., Shafer,
%R. A. \&  Wright, E. L. 
The cosmic microwave background spectrum from the full COBE FIRAS data set.
{\it Astrophys. J.} {\bf 473}, 576-587 (1996).
\v
\item[28.] Hasinger, G. {\it et al.} The ROSAT deep survey. I. X-ray
sources in the Lockman field.
%Burg, R., Giacconi, R., Schmidt, M., Trumper, J. \& Zamorani, G. 
{\it Astron. \& Astrophys.} {\bf 329}, 482-494 (1998).
%\noindent
%26.  Berghoefer, T. W., Bowyer, S., \& Korpela, E. {\it Astrophys.
%J.} {\bf in press} (2000); astro-ph/9912421.\\
%%26. Lieu, R., Bonamente, M. \& Mittaz, J. P. D. {\it Astrophys. J.}
%%{\bf 517}, L91-L95 (1999).\\
%\noindent
%27. Sarazin, C. L. \& Lieu, R. {\it Astrophys. J.} {\bf 494}, L177-L180
%(1998).\\
%\noindent
\v
\item[29.] Kaastra, J. S. High and low energy nonthermal x-ray 
emission from the Abell 2199 cluster of galaxies. {\it et al.} 
{\it Astrophys. J.} {\bf 519}, L119-L122 (1999).
\v
\item[30.] Deiss, B. M., Reich, W., Lesch, H. \& Wielebinsi, R.  The
large-scale structure of the diffuse radio halo of the Coma cluster at
1.4GHz. {\it Astron. \& Astrophys.} {\bf 321}, 55-63 (1997).
\end{itemize}

\normalsize
\vskip 0.2in
\noindent
{ACKNOWLEDGEMENTS.} This work was supported in part by the Israel-US BSF
and by NSF. AL thanks the Weizmann Institute for its kind hospitality
during the course of this work. EW is the incumbent of the Beracha
foundation career development chair.

\newpage

\begin{figure}
\centerline{\psfig{figure=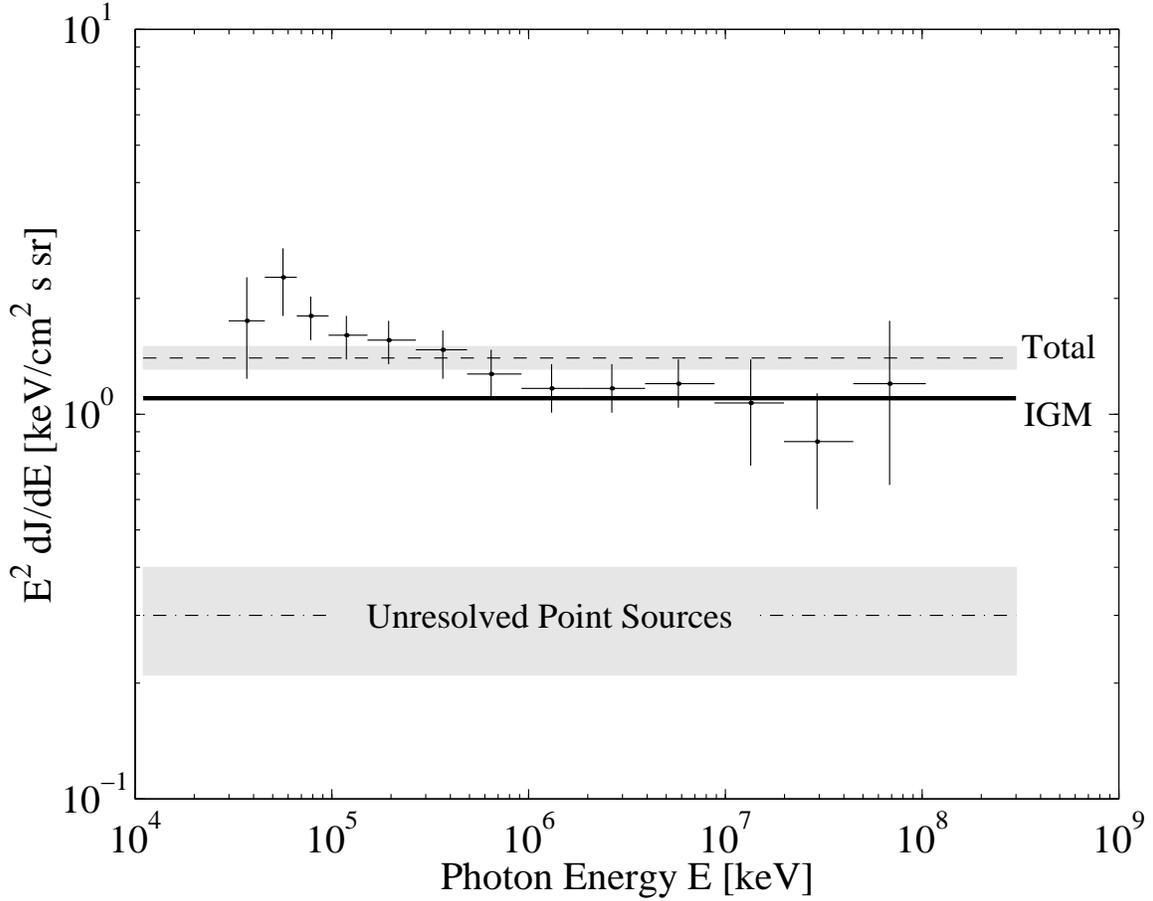,width=6in}}
\caption{Spectrum of the unresolved $\gamma$-ray background.  Points with
error bars are the observed data from EGRET$^1$. The lower shaded region
shows the expected contribution from unresolved point sources, based on
empirical modeling of the luminosity function of blazars$^{2,3}$. The solid
line shows the diffuse emission from intergalactic shocks, according to
Eq.~(1) with $\xi_e=0.05$ and $f_{\rm sh}kT=1$ keV. The upper shaded region
shows the sum of these contributions, which provides an excellent fit to
the data. The blazar contribution was calibrated$^{2,3}$ only by the total
flux in photons with energies $E>10^5$ keV; hence, the small deviations of
some data points from the upper shaded region might be due to uncertainties
in the cumulative blazar spectrum.  Although the background flux predicted
by Eq.~(1) is independent of the magnetic field strength in the
intergalactic shocks, the maximum energy of scattered photons does depend
on the field strength and is given by, $h\nu_{\rm max}\sim
1.2~B_{-7}(kT/{\rm keV})~{\rm TeV}$.  Photons with energies $\ga 1$ TeV
produce an electron-positron pair as they scatter on the infrared
background$^{22,23}$. The pair cools again by scattering microwave photons,
which may in turn produce new pairs.  The energy originally stored in
photons with $h\nu\ga {\rm TeV}$ is spread smoothly over the 100 MeV to TeV
energy range$^{24}$, and might raise the existing flux there. However,
since in our model only a small fraction of the electrons are accelerated
to the energy required for the production of $h\nu\ga {\rm TeV}$ photons,
we expect this effect to be small.  }
\label{fig1}
\end{figure}

\end{document}